\documentclass[conference]{IEEEtran}

\usepackage{epsfig}
\usepackage{multirow}
\usepackage{microtype} 
\usepackage{booktabs}  
\usepackage{url}
\usepackage{paralist}
\usepackage{subfig}
\usepackage{amsmath}
\usepackage{todonotes}
\usepackage[numbers,sort&compress]{natbib}
\usepackage[inline]{enumitem}

\usepackage{color}

\begin{document}
\date{}

\title{Privacy Leakages in Approximate Adders}
\author{\IEEEauthorblockN{Shahrzad Keshavarz, Daniel Holcomb}
	\IEEEauthorblockA{Department of Electrical and Computer Engineering
		\\University of Massachusetts, Amherst, MA, USA 01003
		\\ \{skeshavarz, dholcomb\}@umass.edu }
	\\[-3.0ex]
}

\maketitle
\vspace{-160pt}

\begin{abstract}
Approximate computing has recently emerged as a promising method to meet the low power requirements of digital designs.
The erroneous outputs produced in approximate computing can be partially a function of each chip's process variation. We show that, in such schemes, the erroneous outputs produced on each chip instance can reveal the identity of the chip that performed the computation, possibly jeopardizing user privacy. In this work, we perform simulation experiments on 32-bit Ripple Carry Adders, Carry Lookahead Adders, and Han-Carlson Adders running at over-scaled operating points. Our results show that identification is possible, we contrast the identifiability of each type of adder, and we quantify how success of identification varies with the extent of over-scaling and noise. Our results are the first to show that approximate digital computations may compromise privacy. Designers of future approximate computing systems should be aware of the possible privacy leakages and decide whether mitigation is warranted in their application.

\end{abstract}

\section{Introduction}
\label{sec:introduction}
In recent years, the growing need for energy efficient designs and the emergence of error-tolerant application domains has prompted significant research interest in the area of approximate computing. The basic concept of approximate computing is simple: For many applications such as DSP, data mining and multimedia (audio, video, graphics), a perfect result is usually not necessary. In other words, these classes of applications can tolerate some amount of error. The relaxation of accuracy introduces an amount of design space freedom that can be exploited to reduce power consumption or increase performance.
With predictions of increasing adoption of approximate computing systems in the coming years, designers of approximate computing systems should start considering the associated privacy risks and whether they warrant mitigation.

In this work, we consider how approximate computing can compromise privacy of a device or of a device-bearer.
We assume that an adversary can apply chosen operands to the processor and observe computed, and possibly identifying erronous results.

{\bfseries Contributions: } The specific contributions we make in the paper are as follows:
\begin{itemize}
\itemsep -1pt
\item We show, for the first time, that results from overscaled approximate computations can reveal the identity of the chip that performed the computation.
\item We compare and contrast the identifying ability of the outputs of three popular styles of 32-bit adders.
\end{itemize}

The remainder of this paper is organized as follows: Section~\ref{sec:related_work} provides related work on approximate computing to give context to our contribution. Section~\ref{sec:explanation} explains how approximate computational results can reveal device identity. Section~\ref{sec:methodology} addresses methodology. Section~\ref{sec:evaluation} presents simulation results showing how privacy leakage varies with design and clock frequency. Section~\ref{sec:conclusion} concludes the paper.

\section{Background and Related Work}
\label{sec:related_work}
Approximate circuits exploit the potential error resilience of some classes of applications. This error resilience can have different reasons: \begin{enumerate*}[label=\itshape\alph*\upshape)] \item the data is coming from the real world and therefore, is noisy by nature, \item the algorithm used is self-healing and can attenuate an amount of error, or \item the user of these applications is able to tolerate an amount of error in the result\end{enumerate*} \cite{Venkataramani:2013}. One method of approximate computing is to use deterministic functional approximation, in which a particular Boolean function is replaced by a simpler one that produces similar results at lower complexity~\cite{gupta2011impact, gupta-13}. Because functional approximations compute identical results across all chips, they pose no risk to privacy. 

The computational circuits that are of interest in this work are circuits that use non-deterministic approximations, or what are sometimes denoted timing-based approximations~\cite{Venkatesan:2011jt}. In these approaches, a design is voltage overscaled or frequency overscaled to an operating point where timing constraints may be violated by some circuit paths. At overscaled operating points, the output of a circuit depends not only on inputs, but also on process variation.

\par
Many of the efforts toward approximate computing have focused on adders as ubiquitous basic components of digital systems (e.g.~\cite{gupta-13,gupta2011impact, hegde-01,kedem2011approach,George2006}, among others). More specifically, there has been a lot of research that targets ripple carry adders (RCAs) as an approximate adder of choice because RCAs have a few long paths in the carry chain that are rarely sensitized~\cite{hegde-01}, and this enables a gradual degradation of quality of results when overscaled.
For example, the authors in~\cite{kedem2011approach} have targeted RCAs to reduce the error rate within a fixed energy budget and the authors in~\cite{George2006} proposed a biased voltage scaling for probabilistic RCAs that scales the operating voltage according to the significance of bits.
Because of the focus on adders in previous approximate computing research, we focus our study on adders as well.
\par

Aside from computational blocks in general and adders specifically, there has also been significant interest in approximate memories.
Previous works have proposed DRAM-based approximate memories~\cite{liu-11} with unsafe refresh intervals to save energy, fast but inaccurate writes to multi-level non-volatile storage cells~\cite{sampson-13}, and voltage overscaled SRAM~\cite{esmaeilzadeh-12}.
Recently, one paper has showed that data stored in approximate DRAM can be used as a fingerprint to reveal device identity~\cite{rahmati-15-probable-cause}. To the best of our knowledge, this one previous paper is the only work to explore privacy issues in approximate computing systems, and no previous works at all have studied privacy leakages on the computational (i.e. non-memory) side of approximate computing. 

The use of process variations to identify devices is similar to the idea of a physical unclonable function (PUF) in security. PUFs are circuits designed to extract identifying fingerprints from process variations via timing variations~\cite{gassend-02} or power-up states of SRAM~\cite{guajardo-07, holcomb-09-power-up}.

\section{Identification from Overscaling}
\label{sec:explanation}
Overscaling-based approximate computing relaxes clock period constraints and allows that the long combinational paths of a circuit may not fully propagate within the clock period. In this case, the register at the end of the path may capture intermediate (wrong) results on the clock edge. 
Because of process variation, the critical paths of different chips will have different delays.  For example, recent works report  12\% frequency variation at 1.1V in 45nm technology \cite{7298264} and 30\% for sub-90nm technologies~\cite{Borkar:2003}. The variable path delays will cause different erroneous outputs in approximate computation.
\par

\begin{figure}[htb!]
\setlength{\belowcaptionskip}{-8pt}
\centering
\includegraphics[width=0.7\columnwidth]{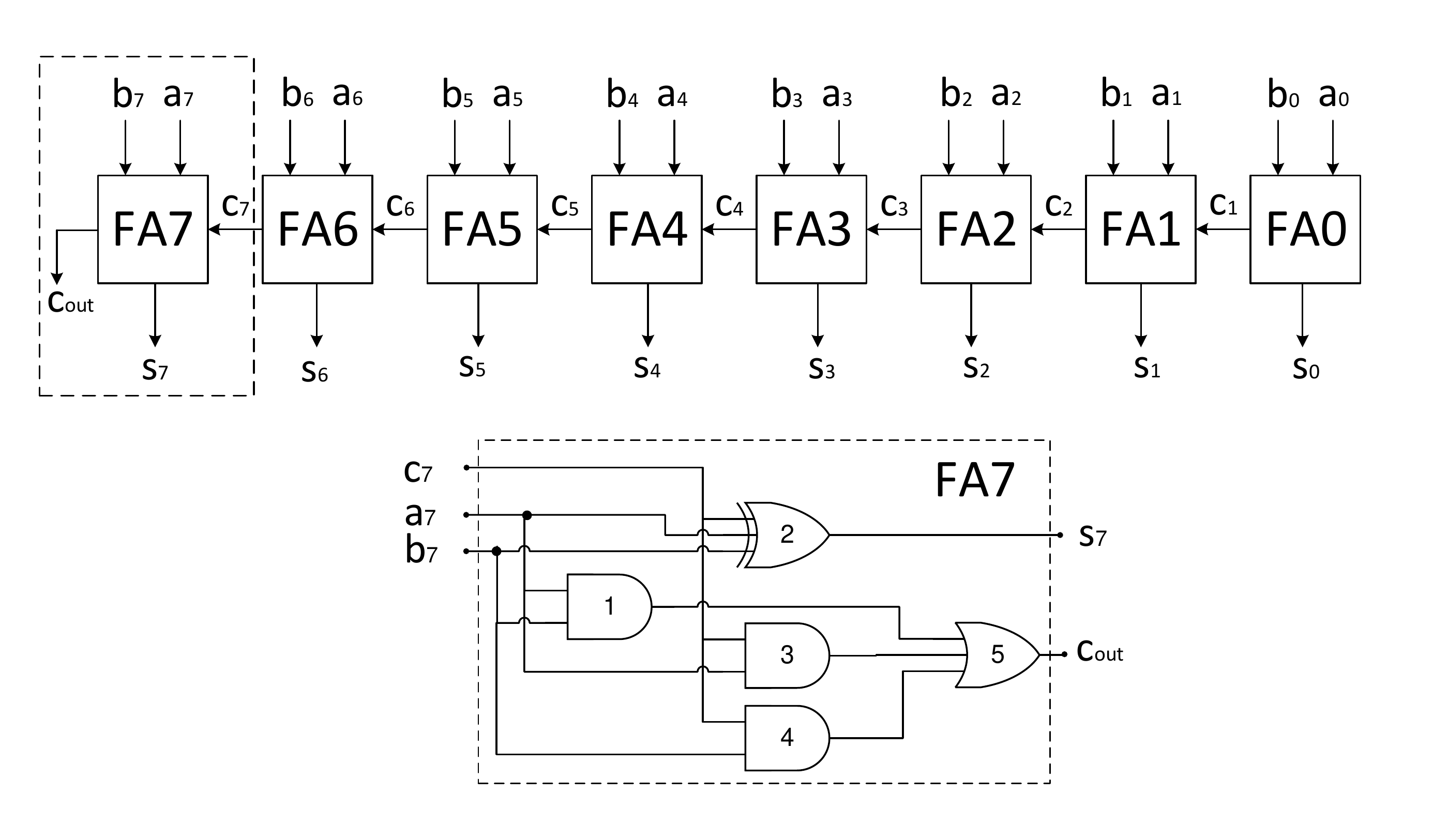}
\caption{An 8-bit ripple carry adder with full-adder blocks.}
\label{fig:FullAdder_complete}
\end{figure}

{\bfseries Example: }
We now give a concrete example to show how gate delays can lead to different results at overscaled operating points. Fig.~\ref{fig:FullAdder_complete} shows an example 8-bit ripple carry adder that has two 8-bit input signals $\{a_7 \dots a_0\}$ and $\{b_7 \dots b_0\}$, and a 9-bit output signal $\{c_{out} s_7 \dots s_0\}$. If $\{a_7 \dots a_0\}=8'b11111111$ and $\{b_7 \dots b_0\}=8'b00000001$, a carry signal has to propagate all the way from FA0 to FA7 in order to generate the correct result. We now focus on what occurs after the carry has propagated through the first seven full adders and signal $c_7$ rises on the input to FA7.
Letting the delay of gate $i$ in FA7 (see Fig.~\ref{fig:FullAdder_complete}) be denoted $d_i$, when the value of $c_7$ rises, the output $s_7$ will fall after time $d_2$. The critical path to $c_{out}$ goes through gates 3 and 5. Therefore, it takes $d_3+d_5$ from the time $c_7$ changes for $c_{out}$ to rise. 

\begin{figure}[htb!]
\centering
\setlength{\belowcaptionskip}{-10pt}
\captionsetup[subfloat]{farskip=20pt,captionskip=0pt}
\subfloat[]{
\label{fig:timing1}
\includegraphics[width=0.25\columnwidth, height=2.5cm]{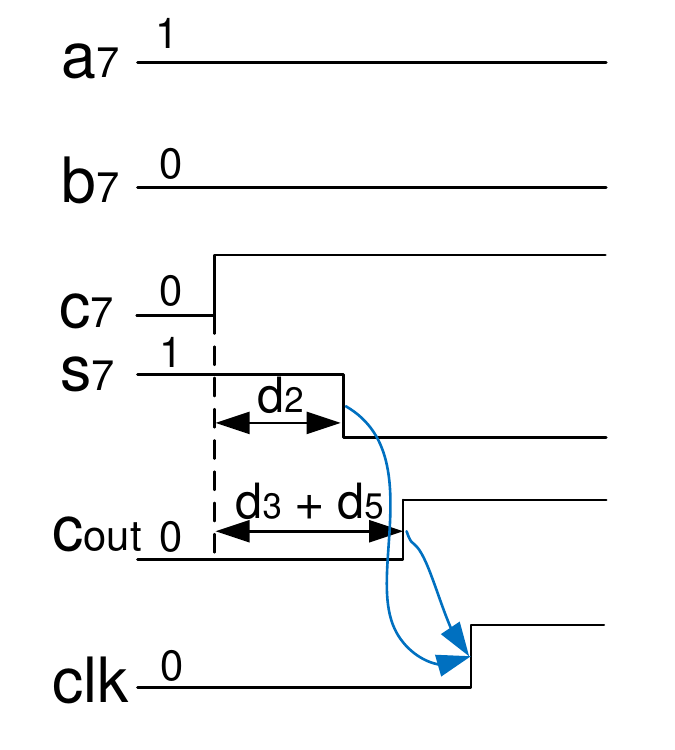}
}
\subfloat[]{
\includegraphics[width=0.25\columnwidth, height=2.5cm]{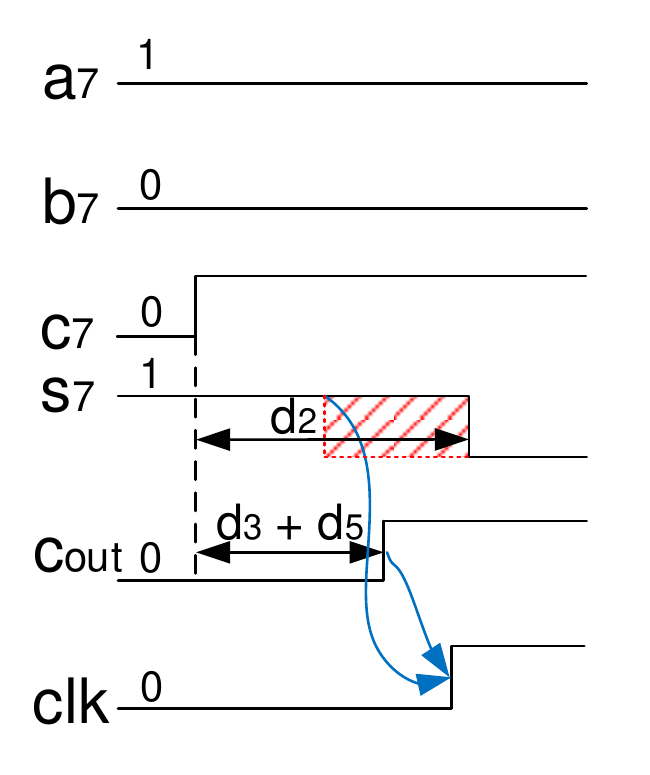}
\label{fig:timing2}
}
\subfloat[]{
\centering
\includegraphics[width=0.25\columnwidth, height=2.5cm]{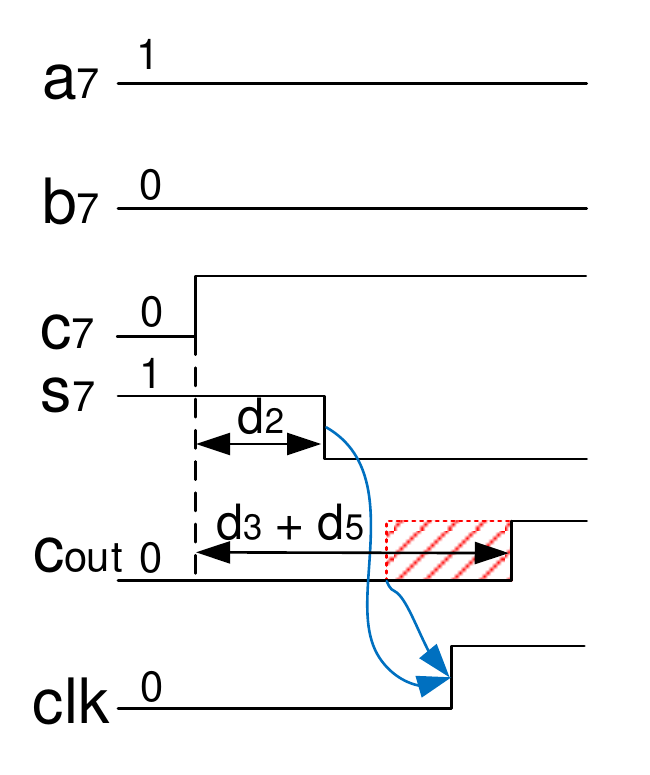}
\label{fig:timing3}
}

\caption{Timing diagram for FA7 of the ripple carry adder depicted in Fig.~\ref{fig:FullAdder_complete}}
\label{fig:timings}
\end{figure}

The value captured on $c_{out}$ and $s_7$ will depend on the delays of the gate instances. In the presence of process variation, some gates might be faster or slower on one chip than another. If all gates are slow relative to the clock period, then the rising transition on $c_7$ may propagate to neither $c_{out}$ nor $s_7$ before the clock edge, and the output will be $c_{out}s_7=01$. If gates 2,3, and 5 are all fast, then the correct value of $c_{out}s_7=10$ will be captured on the clock edge; this is depicted in Fig.~\ref{fig:timing1}. If gate 2 is slow, and gates 3 and 5 are fast, then output $s_7$ will not have fallen before the capturing clock edge, and the captured value will be $c_{out}s_7=11$ (Fig.~\ref{fig:timing2}). If gate 2 is fast and gate 3 or 5 is slow, then output $s_7$ will have fallen but $c_{out}$ will not have risen, and the captured output value will be $c_{out}s_7=00$ (Fig.~\ref{fig:timing3}). This example shows that variations in gate delays can lead to different erroneous outputs in approximate computing; this is the reason that overscaled approximate computing may lead to device identifiability.

\par

{\bfseries Entropy of Input Vectors: } 
Note that for the above example, we only considered a portion of a small circuit. For a large circuit, each individual gate has different delays and many different output results can be generated for some inputs. A good input vector for identification is able to distinguish different chips with different path delays caused by process variation.
When applying random input vectors to a circuit
the majority of vectors will not sensitize long paths and therefore will produce deterministic error-free outputs.
To distinguish the useful vectors from non-useful vectors, we use metric of conditional entropy. When an input vector $a_j$ is applied across a large number of devices at a particular operating point, let the probability of observing output $x_i$ be denoted $Pr(x_i |a_j)$. The entropy associated with the result to input $a_j$ is given by equation~\ref{eq:entropy}. The input vectors with high entropy may be specific to the style of adder and operating point. Although entropy can be estimated from the outputs of adders when viewed as a black box, the entropy associated with different inputs to each adder type implicitly depends on the distribution of path lengths and the path diversity inside of the adder.

\begin{equation}\label{eq:entropy}
H(X|a_j) = - \displaystyle\sum_{i} Pr(x_i|a_j)log_{2}Pr(x_i|a_j)
\end{equation}
\vspace{-10pt}

If an input vector has high entropy on a particular style of adder, it will induce different results for many of the considered chips. 
However in practice, an input vector usually produces the same results for many chips. Furthermore, noise can diminish the usefulness of high-entropy inputs. Nonetheless, entropy is a useful metric that can provide insights about the identifying ability of each adder, as will be discussed in Sec.~\ref{subsec:entropy}.

\section{Methodology}
\label{sec:methodology}
We considere three different 32-bit adders for our evaluations: ripple carry adder (RCA), carry lookahead adder (CLA) and Han-Carlson adder (HCA)~\cite{HanCarlson}. Because our experiments require simulating a large number of vectors on large populations of 32-bit adder circuit instances, using HSPICE simulation alone was found to be impractical in terms of simulation time. Instead, we use HSPICE simulation to extract gate delays and then use timed Verilog simulation with the extracted gate delays to simulate the overall 32-bit circuit. The gate models in HSPICE are 45nm CMOS Predictive Technology Model~\cite{PTM} (PTM) minimum-sized transistors at voltage of 1.0V. Monte-Carlo simulation is performed 100 times across process variations on $V_{th}$ to provide a realistic distribution of pin-to-pin gate delays for each gate type. When creating an instance of the overall adder circuit, we randomly select gate delay instances from the pre-characterized distributions of each gate type. The timed Verilog models of the adders are simulated using Icarus Verilog (iVerilog). 

Changing voltage and changing clock period are two different ways of affecting the same amount of overscaling. In our experiments, we control overscaling by changing the clock period while keeping a set of gate delays extracted at one voltage. We choose this approach because it allows us to dial in a target error rate by performing a binary search on the clock period until hitting the desired error rate.
 We make comparisons across the different styles of 32-bit adders by choosing a clock period for each adder style that realizes equivalent rates of erroneous output. For our 32-bit adders, the clock periods that yield 1\%, 2\% and 5\% erroneous outputs are shown in Tab.~\ref{tab:clock_frequency}. Note that going from 5\% error to 1\% error in an RCA requires increasing the clock period by 127 ps, whereas both CLA and HCA require only a 20 ps increase for the same change in error. This occurs because the RCA has many infrequently sensitized long carry chain paths, whereas HCA is a tree adder with many near-critical paths.


To represent non-idealities in our timing model and to evaluate the robustness of identification, we introduce random noise to our simulations. Each time noise is added to a gate, it is drawn from a normal distribution with 0 mean and standard deviation equal to 10\% of the nominal delay. Noise is uncorrelated across gates, and across vectors, meaning that for each new vector applied to a circuit, the noise offsets are replaced by new values.







\begin{table}[th]
  \centering
  \caption{Clock period used for each adder type to achieve desired error rate.}
{\scriptsize  \begin{tabular}{ c | c | c | c }
           & RCA       & CLA         & HCA \\ \hline
1\% error  & 653 ps      & 345 ps        & 355 ps  \\ \hline
2\% error  & 624 ps      & 340 ps        & 349 ps  \\ \hline
5\% error  & 526 ps      & 325 ps        & 335 ps  \\ \hline
  \end{tabular}
}
  \label{tab:clock_frequency}
\end{table}



\section{Evaluation}
\label{sec:evaluation}
We perform a set of experiments to study the extent to which instances of each adder type can be identified by their outputs. We use these experiments to compare the identifiability of the different adder styles, and the impact of noise.

\subsection{Measuring the Entropy of Vectors}
\label{subsec:entropy}

The first step toward practical chip identification is to pick high entropy vectors. We set the clock period for each adder style to achieve a 1\% error rate and simulate 200,000 random input vectors on 50 instances of RCAs, CLAs and HCAs. We calculate the entropy of each applied vector according to Eq.~\ref{eq:entropy}. The average entropy for RCA, CLA and HCA are 0.01227, 0.01187 and 0.03260, respectively. In the RCA, 97.78\% of all input vectors induce the same result on all 50 chips, while in the HCA, the same number is only 90.73\%. This is because in an RCA, a much higher percentage of vectors cause errors on no chips, or in other words sensitize no paths with delay comparable to or exceeding the clock period. On the other hand, an HCA, which is a tree adder, tends to have a variety of paths with similar nominal delays, and a much lower percentage of vectors are error-free across all chips at the chosen clock period. Another view of this result is as follows: when considering for each adder type the set of random vectors that caused an error on one or more of the 50 instances, we find that each such vector causes errors in about 71\% of RCA chips, versus only 24\% for CLA and 10\% for HCA adders. If each adder type is operated at the same error rate, the error-causing input vectors will be less unique on the RCA. Note however, that non-unique input vectors does not mean that the output vectors are less unique to each chip; instead, it only means that the inputs that {\itshape induce} the erroneous outputs are less unique to each chip.




\subsection{Identification Results} \label{sssec:identification}
\begin{figure}[htb!]
\centering
\setlength{\belowcaptionskip}{0pt}
\captionsetup[subfloat]{farskip=20pt,captionskip=0pt}
\subfloat[Matching distance of ripple carry adder]{
\includegraphics[width=0.8\columnwidth]{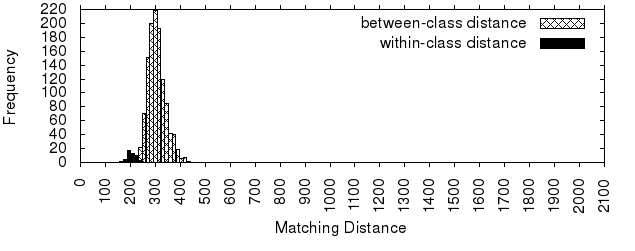}
}\newline
\subfloat[Matching distance of carry lookahead adder]{
\includegraphics[width=0.8\columnwidth]{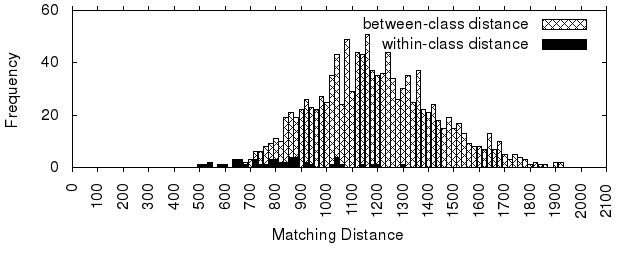}
}\newline
\subfloat[Matching distance of Han-Carlson adder]{
\includegraphics[width=0.8\columnwidth]{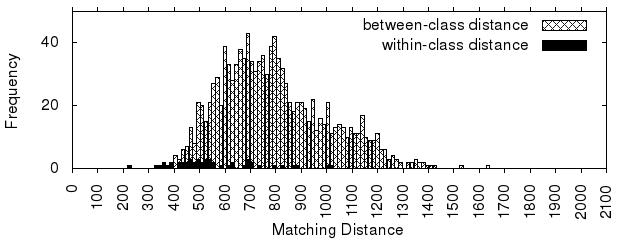}
}
\hfill
\caption{Matching distance based on outputs produced for 40,000 random input vectors for each adder type.}
\label{fig:histogram}
\end{figure}

Next we explore identification of chip instances using their outputs. For this experiment, we simulate 40,000 vectors on 50 instances of each adder type operating at their respective clock periods for 1\% error (Tab.~\ref{tab:clock_frequency}). To measure similarity or lack of similarity between the outputs produced, we use a metric of {\itshape Matching Distance}. The matching distance for any two adders of the same type is the number of outputs that differ when the same (40,000) input vectors are applied. The histograms of between-class and within-class matching distances are shown in Fig.~\ref{fig:histogram}. The within-class bars correspond to the matching distance of two trials of the same chip in the presence of noise (see Sec.~\ref{sec:methodology}), and the between-class bars correspond to pairings of two different chips. When between-class and within-class overlap less, then one can better tell whether two sets of outputs are from the same chip, and can therefore better identify a chip. A chip can always be identified using some matching distance as a decision threshold if the between-class and within-class distances are non-overlapping.

\begin{figure}[htb!]
\centering
\setlength{\belowcaptionskip}{-7pt}
\captionsetup[subfloat]{farskip=20pt,captionskip=2pt}
\includegraphics[width=0.65\columnwidth]{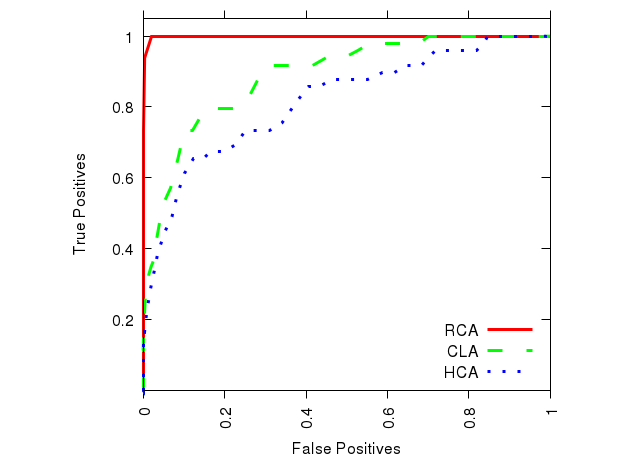}
\caption{ROC curve of three adders when 40,000 vectors are simulated on 50 instances of each adder using a clock period for 1\% error rate. The AUCs for RCA, CLA, and HCA are 0.99, 0.89 and 0.81 respectively.}
\label{fig:ROC_1p}
\end{figure}

{\bfseries ROC Curve:} 
 A Receiver Operating Characteristic (ROC) curve is used to measure the performance of chip identification. Each point of an ROC curve corresponds to a single decision threshold and depicts the trade-off between true positives and false positives at that decision threshold. In an ideal case where between-class and within-class distances are separable, the ROC curve will be a step function~\cite{Jain19991371}, as this would indicate that there exists some decision threshold that can correctly identify all true positives (within-class pairings) without accepting any false positives (between-class pairings). Fig.~\ref{fig:ROC_1p} shows the ROC curve for the three adder styles; the RCA is easily the most identifiable of the three adder styles in this case.

\par

\subsection{Impact of Error Rate}
\begin{figure}[htb!]
	\centering
	\setlength{\abovecaptionskip}{-7pt}
	\captionsetup[subfloat]{farskip=20pt,captionskip=2pt}
	\subfloat[ROC of RCA]{
		\includegraphics[width=0.3\columnwidth]{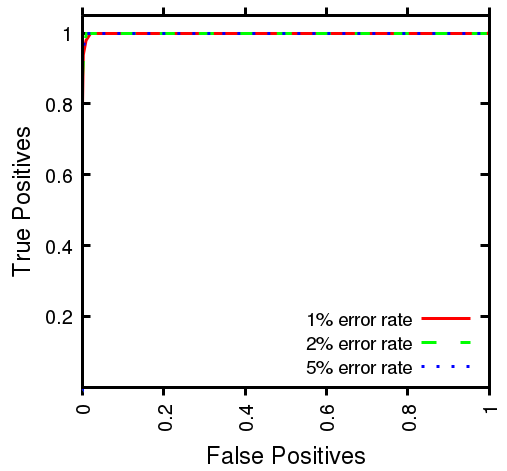}
		\label{fig:ROC_RCA}
	}
	\subfloat[ROC of CLA]{
		\includegraphics[width=0.3\columnwidth]{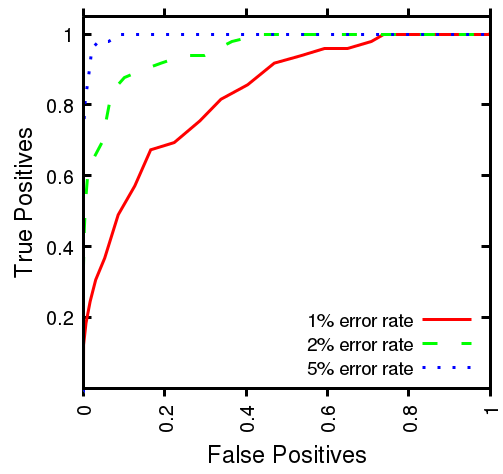}
		\label{fig:ROC_CLA}
	}
\subfloat[ROC of HCA]{
	\includegraphics[width=0.3\columnwidth]{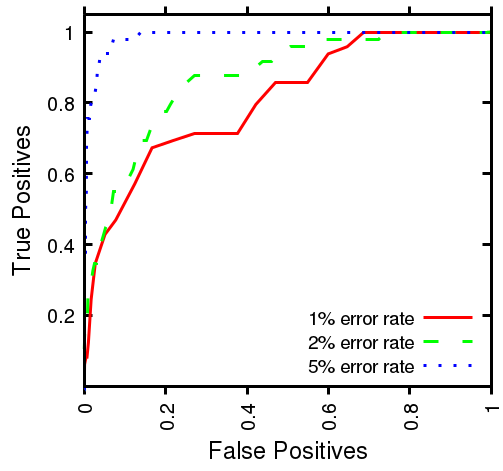}
	
	\label{fig:ROC_HCA}
	}
	\hfill
	\newline \vspace{10pt}
	\caption{ROC curve of each adder type at different error rates}
	\label{fig:ROC_higher}
\end{figure}

There is usually a trade-off between the amount of errors in the outputs and the power/performance improvement in the system. While accepting a higher error rate can be more attractive for efficiency, our results show that a higher error rate can increase identifiability of a circuit. In this experiment we set the clock period such that an average error rate of 1\%, 2\% and 5\% are seen on the output results. The results of this experiment are shown in Fig.~\ref{fig:ROC_higher}.

\section{Conclusion}
\label{sec:conclusion}

This paper has considered, for the first time, the possible privacy implications of voltage-overscaled or frequency-overscaled approximate computations. We perform a large simulation study on three types of adders and show that the ability to provide inputs to a computation unit and observe corresponding outputs can reveal the identity of the approximate computing device that performed the computation. This is a possible privacy risk that designers of future approximate computing systems should consider when evaluating application scenarios.

\par
\textbf{Acknowledgement:} This work has been supported by a grant from the National Science
Foundation (NSF) under award CNS-1563829 and by University of Massachusetts, Amherst.

\thispagestyle{empty}

\footnotesize
\bibliographystyle{acm}
\bibliography{refs}

\begin{thebibliography}{10}

\bibitem{PTM}
{Predictive Technology Model}.
\newblock \url{http://ptm.asu.edu/}.

\bibitem{Borkar:2003}
{\sc Borkar, S., Karnik, T., Narendra, S., Tschanz, J., Keshavarzi, A., and De,
  V.}
\newblock Parameter variations and impact on circuits and microarchitecture.
\newblock In {\em Proceedings of the 40th Annual Design Automation
  Conference\/} (2003), DAC '03, pp.~338--342.

\bibitem{esmaeilzadeh-12}
{\sc Esmaeilzadeh, H., Sampson, A., Ceze, L., and Burger, D.}
\newblock {Architecture support for disciplined approximate programming}.
\newblock In {\em ASPLOS'12: Architectural Support for Programming Languages
  and Operating Systems\/} (Apr. 2012).

\bibitem{gassend-02}
{\sc Gassend, B., Clarke, D., and Van~Dijk, M.}
\newblock {Silicon physical random functions}.
\newblock In {\em Proceedings of the IEEE Computer and Communications
  Society\/} (2002).

\bibitem{George2006}
{\sc George, J., Marr, B., Akgul, B. E.~S., and Palem, K.~V.}
\newblock Probabilistic arithmetic and energy efficient embedded signal
  processing.
\newblock In {\em Proceedings of the 2006 International Conference on
  Compilers, Architecture and Synthesis for Embedded Systems\/} (New York, NY,
  USA, 2006), CASES '06, ACM, pp.~158--168.

\bibitem{guajardo-07}
{\sc Guajardo, J., Kumar, S., Schrijen, G., and Tuyls, P.}
\newblock {{FPGA} intrinsic {PUFs} and their use for {IP} protection}.
\newblock {\em Cryptographic Hardware and Embedded Systems\/} (2007).

\bibitem{gupta2011impact}
{\sc Gupta, V., Mohapatra, D., Park, S.~P., Raghunathan, A., and Roy, K.}
\newblock Impact: imprecise adders for low-power approximate computing.
\newblock In {\em Proceedings of the 17th international symposium on Low-power
  electronics and design\/} (2011), pp.~409--414.

\bibitem{gupta-13}
{\sc Gupta, V., Mohapatra, D., Raghunathan, A., and Roy, K.}
\newblock Low-power digital signal processing using approximate adders.
\newblock {\em Computer-Aided Design of Integrated Circuits and Systems, IEEE
  Transactions on 32}, 1 (Jan 2013), 124--137.

\bibitem{HanCarlson}
{\sc Han, T., and Carlson, D.~A.}
\newblock Fast area-efficient vlsi adders.
\newblock In {\em Computer Arithmetic (ARITH), 1987 IEEE 8th Symposium on\/}
  (May 1987), pp.~49--56.

\bibitem{hegde-01}
{\sc Hegde, R., and Shanbhag, N.~R.}
\newblock {Soft Digital Signal Processing}.
\newblock {\em IEEE Transaction on Very Large Scale Integration (VLSI)
  Systems\/} (2001).

\bibitem{holcomb-09-power-up}
{\sc Holcomb, D.~E., Burleson, W.~P., and Fu, K.}
\newblock Power-up {SRAM} state as an identifying fingerprint and source of
  true random numbers.
\newblock {\em IEEE Transactions on Computers 58}, 9 (Sept. 2009), 1198--1210.

\bibitem{Jain19991371}
{\sc Jain, A.~K., Prabhakar, S., and Chen, S.}
\newblock Combining multiple matchers for a high security fingerprint
  verification system.
\newblock {\em Pattern Recognition Letters 20}, 11-13 (1999), 1371 -- 1379.

\bibitem{kedem2011approach}
{\sc Kedem, Z.~M., Mooney, V.~J., Muntimadugu, K.~K., and Palem, K.~V.}
\newblock An approach to energy-error tradeoffs in approximate ripple carry
  adders.
\newblock In {\em Proceedings of the 17th IEEE/ACM international symposium on
  Low-power electronics and design\/} (2011), pp.~211--216.

\bibitem{liu-11}
{\sc Liu, S., Pattabiraman, K., Moscibroda, T., and Zorn, B.~G.}
\newblock {Flikker: saving DRAM refresh-power through critical data
  partitioning}.
\newblock In {\em Architectural support for programming languages and operating
  systems\/} (June 2011).

\bibitem{rahmati-15-probable-cause}
{\sc Rahmati, A., Hicks, M., Holcomb, D.~E., and Fu, K.}
\newblock Probable cause: The deanonymizing effects of approximate {DRAM}.
\newblock In {\em Proceedings of the 42nd Annual International Symposium on
  Computer Architecture\/} (New York, NY, USA, 2015), ISCA '15, ACM,
  pp.~604--615.

\bibitem{sampson-13}
{\sc Sampson, A., Nelson, J., Strauss, K., and Ceze, L.}
\newblock {Approximate Storage in Solid-State Memories}.
\newblock {\em IEEE Micro\/} (2013).

\bibitem{7298264}
{\sc Tavana, M.~K., Kulkarni, A., Rahimi, A., Mohsenin, T., and Homayoun, H.}
\newblock Energy-efficient mapping of biomedical applications on
  domain-specific accelerator under process variation.
\newblock In {\em Low Power Electronics and Design (ISLPED), 2014 International
  Symposium on\/} (Aug 2014), pp.~275--278.

\bibitem{Venkataramani:2013}
{\sc Venkataramani, S., Chippa, V.~K., Chakradhar, S.~T., Roy, K., and
  Raghunathan, A.}
\newblock Quality programmable vector processors for approximate computing.
\newblock In {\em Proceedings of the 46th International Symposium on
  Microarchitecture\/} (2013), MICRO-46, pp.~1--12.

\bibitem{Venkatesan:2011jt}
{\sc Venkatesan, R., Agarwal, A., Roy, K., and Raghunathan, A.}
\newblock {MACACO: Modeling and analysis of circuits for approximate
  computing}.
\newblock In {\em International Conference on Computer-Aided Design (ICCAD)\/}
  (2011), pp.~667--673.

\end{thebibliography}

\end{document}